\documentclass{article}
\usepackage[utf8]{inputenc}
\usepackage[T1]{fontenc}
\usepackage{graphicx}
\usepackage{longtable}
\usepackage{rotating}
\usepackage{amsmath}
\usepackage{amssymb}
\usepackage{hyperref}
\usepackage{color}
\usepackage{microtype}
\usepackage{lmodern}
\usepackage{subcaption}

\newcommand{\p}[1]{\mathbf{#1}}
\newcommand{\intf}[5]{\int_{#1}^{#2}d x_i f_{\X_i}(x_i)\int_{#3}^{#4}d y_i f_{\Y_i}(y_i)\,{#5}}
\newcommand{\erf}{\mathrm{erf}}
\newcommand{\X}{\mathcal{X}}
\newcommand{\Y}{\mathcal{Y}}
\newcommand{\J}{\mathcal{J}}
\newcommand{\cS}{\mathcal{S}}

\title{An Analytical Approach to \\ 
the Jaccard Similarity Index}
\author{Gonzalo Travieso \and Alexandre Benatti \and Luciano da F. Costa \\
\small São Carlos Institute of Physics, University of São Paulo, Brazil}
\date{October, 2024}

\begin{document}
\maketitle

\begin{abstract}
The Jaccard similarity index has often been employed in science and technology as a means to quantify the similarity between two sets. When modified to operate on real-valued values, the Jaccard similarity index can be applied to compare vectors, an operation which plays a central role in visualization, classification, and modeling. The present work aims at developing an analytical approach for estimating the probability density of the Jaccard similarity values as implied by set of data elements characterized by specific statistical densities, with emphasis on the uniform and normal cases. Several theoretical and practical situations can benefit directly from such an approach, as it allows several of the properties of the similarity comparisons among a given dataset to be better understood and anticipated. Situations in which the described approach can be applied include the estimation and visualization of data interrelationships in terms of similarity networks, as well as diverse problems in data analysis, pattern recognition and scientific modeling. In addition to presenting the analytical developments and results, examples are also provided in order to illustrate the potential of the approach. The work also includes extension of the reported developments to modifications of the Jaccard index intended for regularization and control of the sharpness of the implemented comparisons.
\end{abstract}

\section{Introduction}\label{sec:introduction}

Science and technology are largely based on quantifying properties of abstract or real-world entities and systems in terms of respective \emph{measurements} (or features), as this contributes to the objectivity of representation, characterization, and modeling approaches. Another key aspect of science and technology which is not so often realized concerns the extensive need of means for \emph{relating} and \emph{comparing} quantitative representations. An example of the central role played by comparisons include the validation of a model, in which simulated predictions need to be contrasted against respective experimental data. As another example, we have taking decisions about a given problem while quantitatively comparing its properties with previously known cases, so that a decision can eventually be reached while taking into account the most similar already known situations. In pattern recognition (e.g.~\cite{fukunaga1993,duda2000pattern,theodoridis2006pattern}), comparisons are frequently employed as a means of identifying relationships between data elements and respective clusters. An ample range of activities in science and technology involve comparisons. For all the above reasons, the activity of comparing two entities represented in terms of respective features constitutes an issue deserving special attention.

Reflecting the important role of comparisons, several means of implementation have been developed. These include distances (e.g.~\cite{strang2022}), correlations and statistical joint variation (e.g.~\cite{johnsonwhichern}), inner products~\cite{strang2022}, as well as similarity indices (e.g.~\cite{wolda1981,sorensen1948,Lieve1989,kabir2017,steinley2021}), each with its specific properties that can be intrinsically more appropriate for given applications. Among several possible interesting similarity indices (e.g.~\cite{costasimil}), the \emph{Jaccard similarity} (e.g.~\cite{jaccard1901,Jac:wiki,costafurther}) has received substantial attention, especially as a means to compare two sets. Basically, the Jaccard index can be calculated by dividing the cardinality of the intersection between the two compared sets by the cardinality of their respective union. The basic rationale underlying this index is that two similar sets will lead to similar unions and intersections. This index provides as result a dimensionless scalar values comprised within the interval $[0,1]$.

The Jaccard similarity index has been modified in order to compare two non-zero real-valued vectors composed of non-negative entries (e.g.~\cite{costafurther,costaintegrate}), and then to generic real-valued vectors, as well as potentially any type of mathematical structures including functions (e.g.~\cite{costafurther, costasimil}). As a consequence of the non-linear nature (use of maximum and minimum) of this index, it tends to perform particularly sharper comparisons than alternative approaches including the cosine similarity, inner products, and distances. It is also possible to consider taking the obtained similarity values to a power of $D$ as a means of enhancing in a controlled manner the sharpness of the implemented comparisons~\cite{costafurther}. Interestingly, the higher the value of $D$, the sharper the comparison becomes, thus providing an effective means of implementing, in controlled manner, comparisons between similar entities. By combining the Jaccard and interiority index (also overlap, e.g.~\cite{Kavitha}), it has also been possible to obtain another index, namely the coincidence similarity index~\cite{costafurther,costaintegrate,costamneurons}, which further enhances the comparisons. In addition, the proportional nature of the Jaccard index has been verified to provide an interesting manner to approach data classification involving right skewed features~\cite{benatti2024supervised}.

By allowing comparisons to be implemented in terms of similarities, instead of distances, it becomes possible to estimate \emph{similarity networks} associated to a dataset (e.g.~\cite{costacoinc}). More specifically, given a set of entities (data elements) of interest represented in terms of respective properties or \emph{features}, a similarity network can be obtained by representing each of the entities as a node, while links between pairs of nodes can then be assigned weights corresponding to similarity between those pairs of nodes. Similarity networks can provide interesting information about the original dataset, especially concerning the \emph{interrelationship} between the original entities, allowing the identification of possible \emph{heterogeneity} among the interconnections, as well as the eventual existence of modules (groups of nodes more intensely interconnected).

Oftentimes, when applying the Jaccard similarity index to a given dataset, it becomes of special interest to estimate the distribution of similarity values which can be respectively observed. This constitutes a not so straightforward problem, especially when the data elements are characterized in terms of many features, and also as a consequence of the non-linear nature of the multiset Jaccard index. Having means for analytically estimating the distribution of Jaccard similarity values in multivariate cases modeled by respective types of densities (e.g.~uniform, normal, etc.) can help better understanding several of similarity-related properties of the studied datasets, including their heterogeneity, skewness, magnitude variations, as well as eventual presence of multimodality. In addition, the ability to estimate the density of Jaccard similarity values respectively to a given dataset characterized by a respective density can also assist and complement in a significant manner the task of obtaining similarity networks from a given dataset by allowing some of their respective topological properties to be anticipated. Indeed, analytical means for estimating the statistical distribution of Jaccard similarity index are intrinsically interesting from the perspective of most of the situations in which this index has been applied.

The present work has been aimed at developing an analytical approach to estimating the distribution of Jaccard similarity values obtained from data elements characterized by multivariate real-valued features described by respective density functions, with emphasis on uniform and normal densities.

This work starts by presenting assumptions about the considered data distributions so as to allow the central limit theorem (e.g.~\cite{fischer2011}) to be applied. Developments are then described leading to analytical expressions of the average and variance of the sought distribution, as well as an expression from which this distribution can be estimated. Illustration of the potential of the reported developments are also provided in terms of density plots and similarity networks relatively to uniform and normal data distributions. The extension of the analytical approaches to modified Jaccard indices aimed at regularizing and controlling the sharpness of the implemented comparisons have also be included and illustrated in this work.

\section{Jaccard similarity}
\label{sec:jaccard}

Consider two points, \(\p{x}\) and \(\p{y}\), in a multivariate space with \(n\) dimensions, independently drawn from random variables \(\X\) and \(\Y\) with probability densities given by \(f_\X(\p{x})\) and \(f_\Y(\p{y})\), respectively.
Their Jaccard similarity generalized to real-valued vectors can be expressed~\cite{costasimil,costaintegrate} as
\begin{equation}\label{eq:def}
J(\p{x}, \p{y}) =
    \frac{\sum_{i=1}^{n}\min(x_i^P, y_i^P)+\min(|x_i^N|,|y_i^N|)}
         {\sum_{i=1}^{n}\max(x_i^P, y_i^P)+\max(|x_i^N|,|y_i^N|)},
\end{equation}
where \(x_i^P = x_i\) if \(x_i\ge0\) and \(x_i^P=0\) otherwise, while \(x_i^N = x_i\) if \(x_i\le0\) and \(x_i^N=0\) otherwise, and similarly for \(y_i^P\) and \(y_i^N\).

The Jaccard similarity can be understood as a random variable \(\J\) of which it is of interest to compute the distribution.
Consider first the probability that the similarity is less than or equal to a given value \(J\), that is, the probability of
\begin{multline}\label{eq:sum}
  \sum_{i=1}^{n} \left[\left(\min(x_i^P, y_i^P)+\min(|x_i^N|,|y_i^N|)\right) - \right.\\
    \left.-J \left(\max(x_i^P, y_i^P)+\max(|x_i^N|,|y_i^N|)\right)\right]\le0.
\end{multline}

To carry out this calculation the following simplifying assumptions are introduced and adopted:
\begin{enumerate}
\item The different dimensions \(i\) are statistically independent.
  That is, \(f_\X(\p{x})=\prod_{i=1}^n f_{\X_i}(x_i)\), and similarly for \(f_{\Y}\).
\item The number of dimensions \(n\) is sufficiently large.
\end{enumerate}

With these assumptions, the sum in Eq~\eqref{eq:sum} can be approximated using the central limit theorem.
For this the mean \(\mu_i(J)\) and variance \(\sigma_i^2(J)\) for each dimension \(i\) are needed.
The sum will be approximated by a normal distribution with mean
\begin{equation}\label{eq:mu}
  \mu(J) = \sum_{i=1}^{n} \mu_i(J)
\end{equation}
and variance
\begin{equation}\label{eq:sigma2}
  \sigma^2(J) = \sum_{i=1}^{n}\sigma_i^2(J).
\end{equation}

The probability of a sample of this normally distributed variable having a value less than or equal to zero is an approximation to the cumulative probability of \(\J\). This probability is given by
\begin{equation}
  \label{eq:cdfnorm}
  \frac{1}{2}\left[1+\erf\left(- \frac{\mu(J)}{\sqrt{2\sigma^2(J)}}\right)\right].
\end{equation}
An approximation to the probability density function (PDF) of \(\J\) is provided therefore by the derivative of this expression with respect to \(J\),
\begin{equation}
  \label{eq:deriv}
  f_\J(J)=\frac{\mu\frac{d\sigma^2}{dJ}-2\frac{d\mu}{dJ}\sigma^2}{2\sqrt{2\pi}(\sigma^2)^{3/2}}\exp\left(-\frac{\mu^2}{2\sigma^2}\right).
\end{equation}

To compute \(\mu_i(J)\) and \(\sigma^2_i(J)\), six possible cases for the relative values of \(x_i\) and \(y_i\) need to be considered, as shown in Table~\ref{tab:cases}. The contribution of dimension $i$ to the sum \eqref{eq:sum} is therefore
\begin{align}
  \label{eq:meani}
  \mu_i =  & \intf{-\infty}{0}{-\infty}{x_i}{(-x_i + J y_i)} + \nonumber \\
         +  & \intf{-\infty}{0}{x_i}{0}{(-y_i + J x_i)} + \nonumber \\
         +  & \intf{-\infty}{0}{0}{\infty}{(-J y_i + J x_i)} + \nonumber \\
         +  & \intf{0}{\infty}{-\infty}{0}{(-J x_i + J y_i)} + \nonumber \\
         +  & \intf{0}{\infty}{0}{x_i}{(y_i - J x_i)} + \nonumber \\
         +  & \intf{0}{\infty}{x_i}{\infty}{(x_i - J y_i)},
\end{align}
and the variance can be estimated as

\begin{align}
  \label{eq:vari}
  s_i = & \intf{-\infty}{0}{-\infty}{x_i}{(-x_i + J y_i)^2} + \nonumber \\
         + & \intf{-\infty}{0}{x_i}{0}{(-y_i + J x_i)^2} + \nonumber \\
         + & \intf{-\infty}{0}{0}{\infty}{(-J y_i + J x_i)^2} + \nonumber \\
         + & \intf{0}{\infty}{-\infty}{0}{(-J x_i + J y_i)^2} + \nonumber \\
         + & \intf{0}{\infty}{0}{x_i}{(y_i - J x_i)^2} + \nonumber \\
         + & \intf{0}{\infty}{x_i}{\infty}{(x_i - J y_i)^2}, \nonumber \\
  \sigma_i^2 = & s_i - \mu_i^2.
\end{align}

\begin{sidewaystable}
\begin{center}
\begin{tabular}{l|ccccccccc}
Case & \(x_i^P\) & \(|x_i^N|\) & \(y_i^P\) & \(|y_i^N|\) & \(\min(x_i^P, y_i^P)\) & \(\min(|x_i^N|, |y_i^N|)\)& \(\max(x_i^P, y_i^P)\) & \(\max(|x_i^N|, |y_i^N|)\) & Factor\\[0pt]
\hline
\(x < 0, \quad y\le x\)   & 0       & \(-x_i\) & 0       & \(-y_i\)  & 0       & \(-x_i\)  & 0       & $-y_i$ & \(-x_i + J y_i\)\\[0pt]
\(x<0, \quad x < y < 0\)  & 0       & \(-x_i\) & 0       & \(-y_i\)  & 0       & \(-y_i\)  & 0       & $-x_i$ & \(-y_i + J x_i\)\\[0pt]
\(x < 0, \quad y \ge 0\)  & 0       & \(-x_i\) & \(y_i\) & 0         & 0       & 0         & \(y_i\) & $-x_i$ & \(-J y_i + J x_i\)\\[0pt]
\(x\ge0, \quad y<0\)      & \(x_i\) & 0        & 0       & \(-y_i\)  & 0       & 0         & \(x_i\) & $-y_i$ & \(-J x_i + J y_i\)\\[0pt]
\(x\ge0, \quad 0<y\le x\) & \(x_i\) & 0        & \(y_i\) & 0         & \(y_i\) & 0         & \(x_i\) & 0     &  \(y_i - J x_i\)\\[0pt]
\(x\ge0, \quad y>x\)      & \(x_i\) & 0        & \(y_i\) & 0         & \(x_i\) & 0         & \(y_i\) & 0     &  \(x_i - J y_i\)\\[0pt]
\end{tabular}
\end{center}
\caption{Computation of \(\min(x_i^P, y_i^P)\), \(\min(|x_i^N|, |y_i^N|)\), \(\max(x_i^P, y_i^P)\), and \(\max(|x_i^N|, |y_i^N|)\) for the six different cases.}\label{tab:cases} 
\end{sidewaystable}

Given the densities \(f_{\X_i}\) and \(f_{\Y_i}\) for each coordinate, it is possible to use these expressions to compute, either analytically or numerically, the PDF of the Jaccard similarity using Eq.~\eqref{eq:deriv}.
The following section presents some concrete examples.

\section{Some examples}
\label{sec:some-examples}

In this section, we illustrate the application of the above described analytical resources to some types of data characterized by uniform and normal densities.

A first simple case with different distributions for \(\X\) and \(\Y\) is
\begin{align}
  & \X_i \sim U(-1/4, 3/4), \quad \Y_i \sim U(-3/4, 1/4), & i=1,\ldots,s, \nonumber\\
  & \X_i \sim U(-1/2, 1/2), \quad \Y_i \sim U(-1/2, 1/2), & i=s+1,\ldots,n,
\label{eq:fxfyuni}
\end{align}
where \(U(a, b)\) denotes the uniform distribution between \(a\) and \(b\).
As the last \(n-s\) dimensions are identically distributed for \(\X\) and \(\Y\), the parameter \(s\) controls the level of superimposition between the two random variables.
But note that even in the initial \(s\) dimensions, there is already a significant amount of superimposition regarding half the ranges in each dimension.

Considering these distributions in Eqs.~\eqref{eq:meani} and~\eqref{eq:vari} results in \(\mu_i=(1-14J)/24\) and \(\sigma^2_i=(11+31J+242J^2)/2304\) for \(i=1,\ldots,s\) and \(\mu_i=(1-5J)/12\) and \(\sigma^2_i=(2+J+5J^2)/144\) for \(i=s+1,\ldots,n\), or
\begin{align*}
\mu &= \frac{2n - s - (10n + 4s)J}{24},\\
\sigma^2 & = \frac{32 n - 21 s + (16 n + 15 s) J + (80 n + 162 s) J^2}{2304}.
\end{align*}

Therefore, the PDF of \(\J\) can be estimated as
\begin{multline}
  \label{eq:pdfdifferent}
  f_\J(J) = \frac{3}{\sqrt{2\pi}}
  \frac{224 n^2 - 50 n s - 61 s^2 + (160 n^2 + 234 n s - 88 s^2) J}
       {[32 n - 21 s + (16 n + 15 s) J + (80 n + 162 s) J^2]^{3/2}}\times\\
  \times \exp\left(-\frac{2 [-2 n + s + (10 n + 4 s) J]^2}
                  {32 n - 21 s + (16 n + 15 s) J + (80 n + 162 s) J^2}
      \right).
\end{multline}

\begin{figure}[t]
  \centering
  \includegraphics[width=0.48\textwidth]{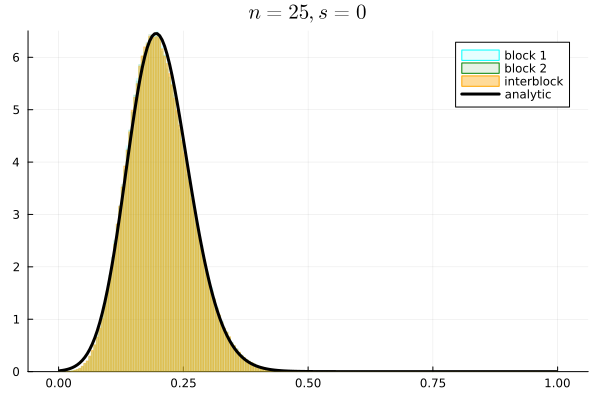}
  \includegraphics[width=0.48\textwidth]{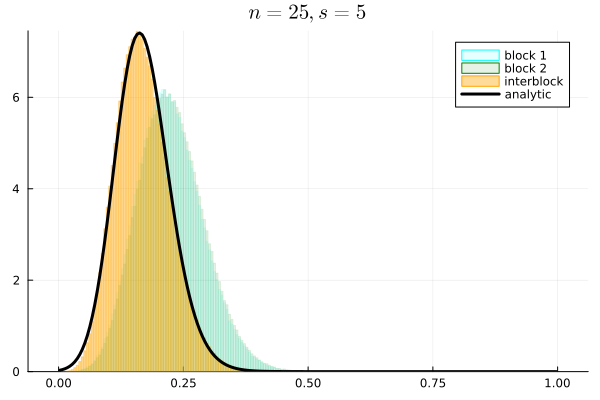}\\
  \includegraphics[width=0.48\textwidth]{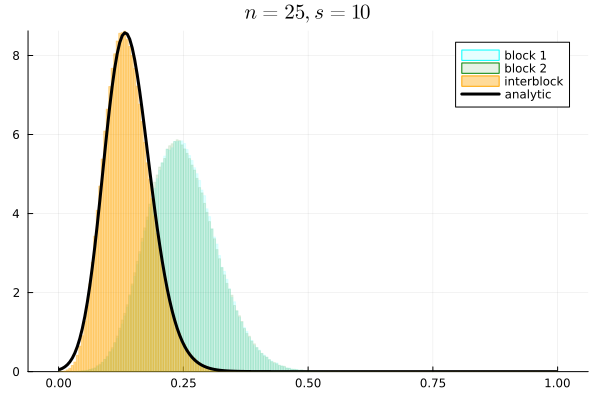}
  \includegraphics[width=0.48\textwidth]{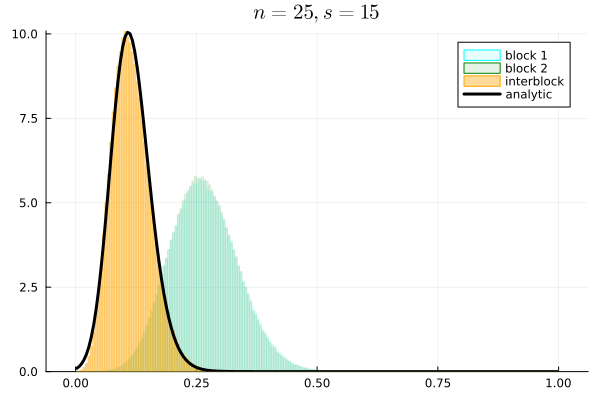}\\
  \includegraphics[width=0.48\textwidth]{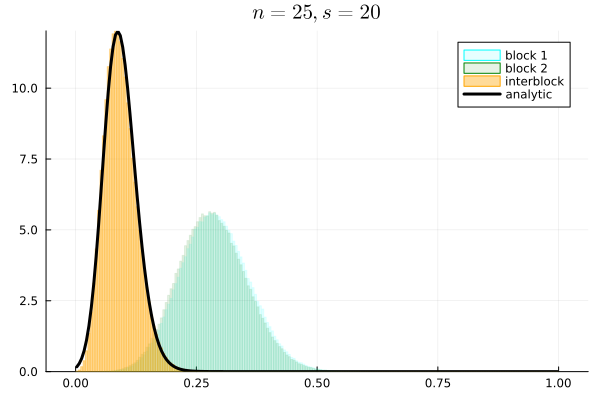}
  \includegraphics[width=0.48\textwidth]{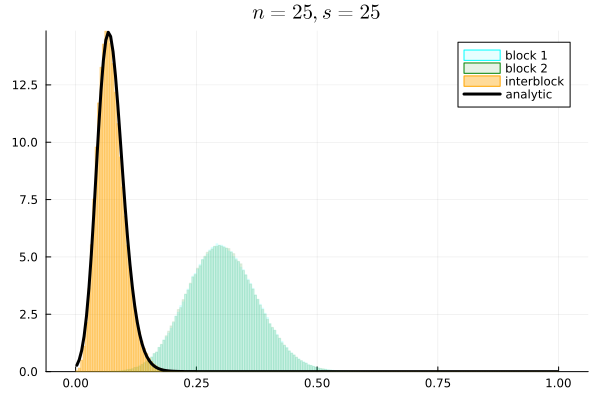}
  \caption{
    Comparison of Eq.~\eqref{eq:pdfdifferent} with sample data.
    A total of 1000 data points each are generated for $\X$ (block 1) and $\Y$ (block 2) described by Eq.~\eqref{eq:fxfyuni}, with $n=25$ and $s=0, 5, 10, 15, 20, 25$.
  }
  \label{fig:diffcomp}
\end{figure}

This expression is compared with the similarities computed by randomly generated data in Figure~\ref{fig:diffcomp}. For reference, the distribution of \(\J\) inside the blocks, that is, of data generated by \(f_\X\) with itself and the same for \(f_\Y\) is also included.

It can be seen that Eq.~\eqref{eq:pdfdifferent} gives a good approximation to the distribution of values of \(\J\).
Some discrepancy can be observed near \(J=0\), as the normal approximation used to reach Eq.~\eqref{eq:cdfnorm} yields finite probabilities outside of the region \([0, 1]\) where \(J\) is confined. Observe that the block densities are necessarily identical as a consequence of the intrinsic symmetry of the original dataset.

Several interesting insights can be obtained from the results shown in Figure~\ref{fig:diffcomp}, especially regarding the manner in which the progressive increase of the parameter \(s\) leads to and increased separation between the densities of similarities between the two block and inside the blocks. Also associated with an increase in \(s\) is a reduction of the values of similarities between blocks and a closer concentration of values near the peak of the density (reduced variability in the values of \(J\)).

\begin{figure}[t]
  \centering
  \includegraphics[width=0.5\textwidth]{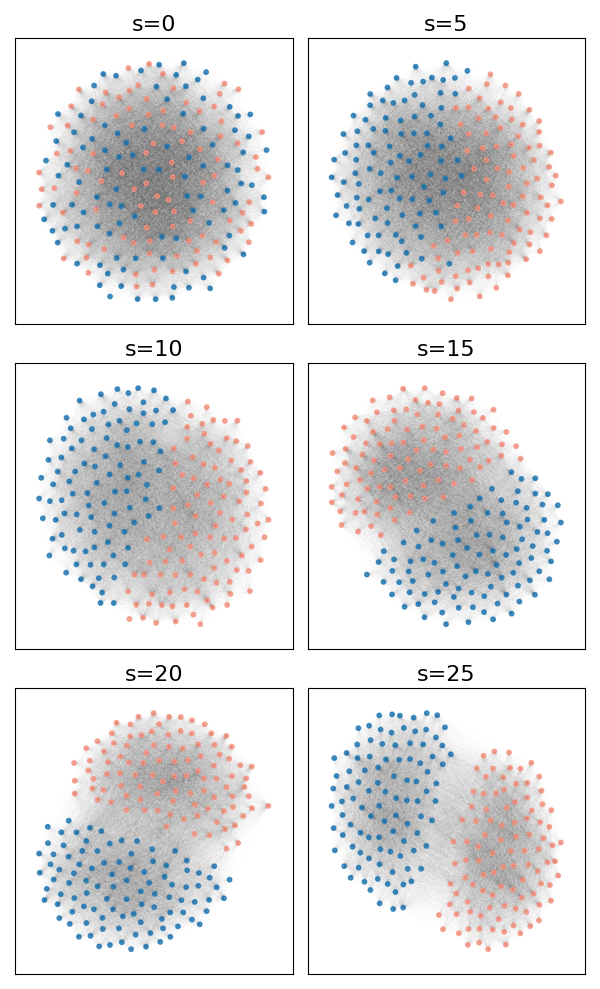}
  \caption{Illustration of the cases considered in Fig.~\ref{fig:diffcomp} in terms of Jaccard similarity networks obtained by using the respective parameters. Each of the two blocks of 100 points are represented by distinct colors. A progressive separation of the two groups as $s$ increases can be observed.}
  \label{fig:uniform_s}
\end{figure}

Figure~\ref{fig:uniform_s} presents an illustration of the cases in Figure~\ref{fig:diffcomp} in terms of respective Jaccard similarity networks, where each point is represented by a respective node, and the weight of the links correspond to the respective Jaccard similarities. The points belonging to the two considered blocks are shown in distinct colors. It can be readily verified that the two groups of points tend to progressively separate one another as the parameter $s$ is increased, defining two respective modules or communities. All networks have been visualized by using the Fruchterman-Reingold methodology~\cite{fruchterman1991}.

As another example, consider
\begin{align}
  \label{eq:fxfynorm}
  & \X_i \sim N(1/2, 1/3), \quad \Y_i \sim N(-1/2, 1/3), & i=1,\ldots,s \nonumber\\
  & \X_i \sim N(0, 1/3), \quad \Y_i \sim N(0, 1/3), & i=s+1,\ldots,n,
\end{align}
where \(N(m, d)\) denotes the normal distribution with mean \(m\) and standard deviation \(d\).

Again, \(s\) controls the amount of superimposition of the densities of the two random variables.

Inserting in Eq.~\eqref{eq:meani} and integrating numerically leads to 
\begin{align*}
  \mu_i & = 0.016664 - 1.022412 J & i=1,\ldots,s\\
  \mu_i & = 0.077898 - 0.454025 J & i=s+1,\ldots,n.\\
\end{align*}
Similarly, for Eq.\eqref{eq:vari}
\begin{align*}
  \sigma^2_i & = 0.003617 + 0.014154 J + 0.192921 J^2 & i=1,\ldots,s\\
  \sigma^2_i & = 0.014120 + 0.066632 J^2 & i=s+1,\ldots,n.\\
\end{align*}
Therefore
\begin{align*}
  \mu & = 0.077898 n - 0.061235 s - (0.454025 n + 0.568387 s) J\\
  \sigma^2 & = 0.014120 n - 0.010503 s + 0.014154 s J + (0.66632 n + 0.126289 s) J^2.
\end{align*}

\begin{figure}[t]
  \centering
  \includegraphics[width=0.48\textwidth]{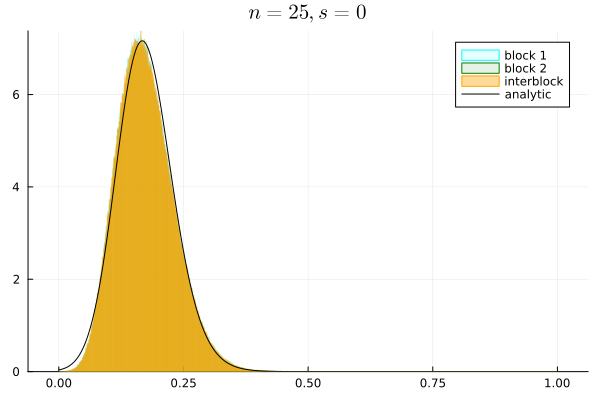}
  \includegraphics[width=0.48\textwidth]{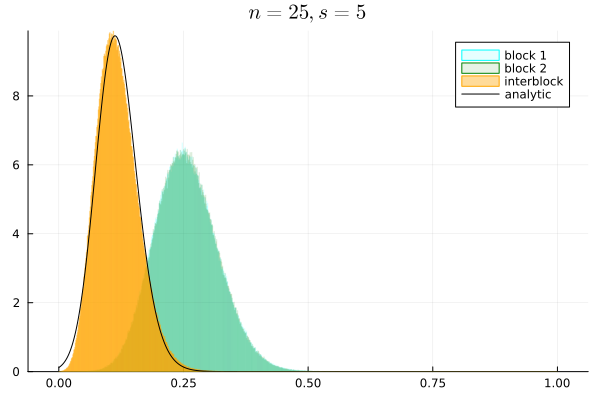}\\
  \includegraphics[width=0.48\textwidth]{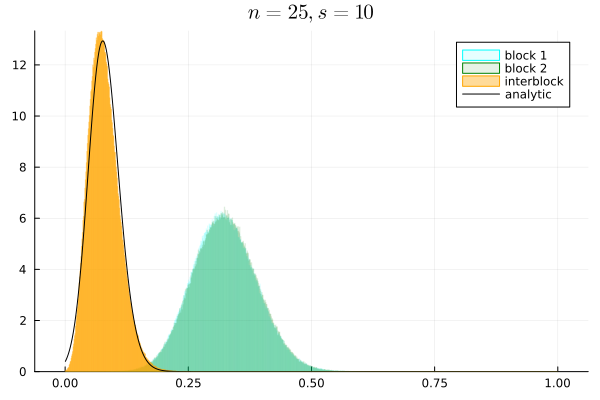}
  \includegraphics[width=0.48\textwidth]{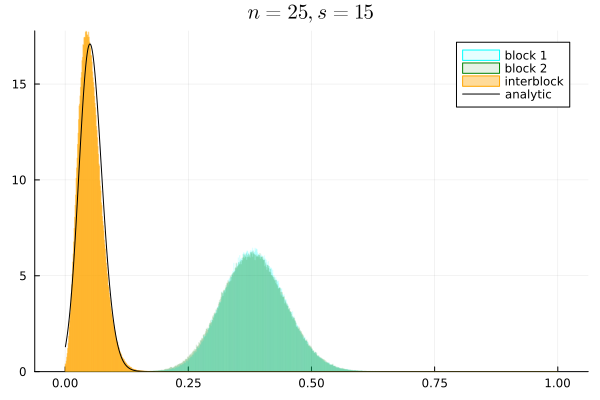}\\
  \includegraphics[width=0.48\textwidth]{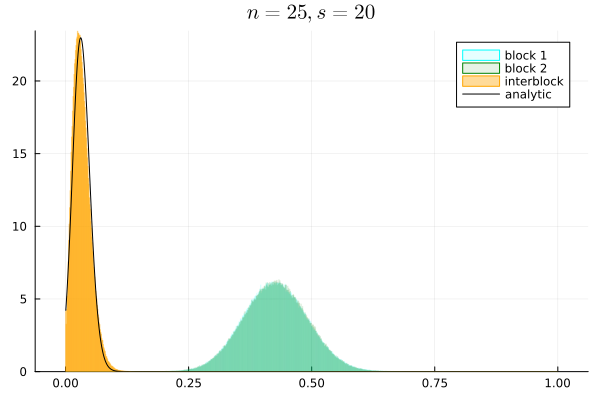}
  \includegraphics[width=0.48\textwidth]{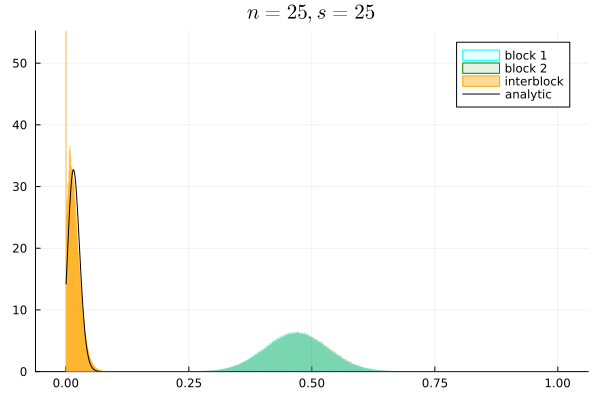}
  \caption{
    Comparison of the expression for of the distribution of values of $J$ for the case of Eq.~\eqref{eq:fxfynorm} with sample data.
    A total of 1000 data points each are generated for $\X$ (block 1) and $\Y$ (block 2), with $n=25$ and $s=0, 5, 10, 15, 20, 25$.
  }
  \label{fig:diffcompnorm}
\end{figure}

\begin{figure}[t]
  \centering
  \includegraphics[width=0.48\textwidth]{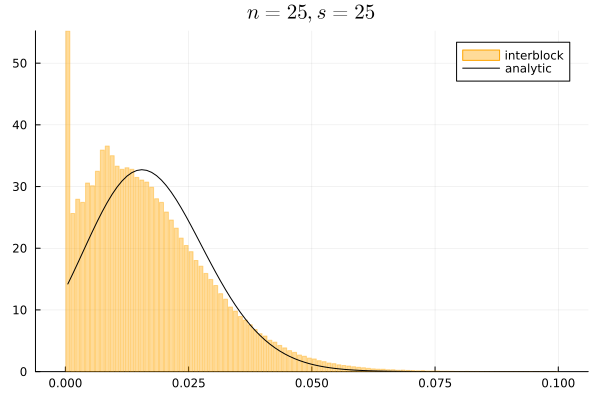}
  \includegraphics[width=0.48\textwidth]{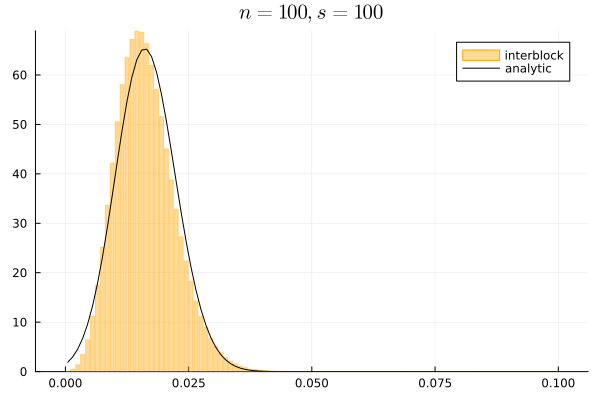}
  \caption{
    Effect of increased dimensionality in the approximation. Comparison of $n=25, s=25$ with $n=100, s=100$.}
  \label{fig:largen}
\end{figure}

Inserting in Eq.~\eqref{eq:deriv} yields the expression for \(f_J(J)\), which is omitted here. A comparison with numeric data is found in Fig.~\ref{fig:diffcompnorm}.
We see increased discrepancies near $J=0$ in comparison to Fig.~\ref{fig:diffcomp}.
These discrepancies which are due to our normal approximation, decrease when the dimensionality of the data is larger, as illustrated in Fig.~\ref{fig:largen}.

As in the case of the previous examples, involving uniform densities, the block densities progressively separated from the interblock density as \(s\) increased. In this case, the separation of densities is larger and the interblock density is closer to zero and more sharply peaked. The adopted analytical approach allows a precise quantitative understanding of the involved dynamics.

\begin{figure}[t]
  \centering
  \includegraphics[width=0.5\textwidth]{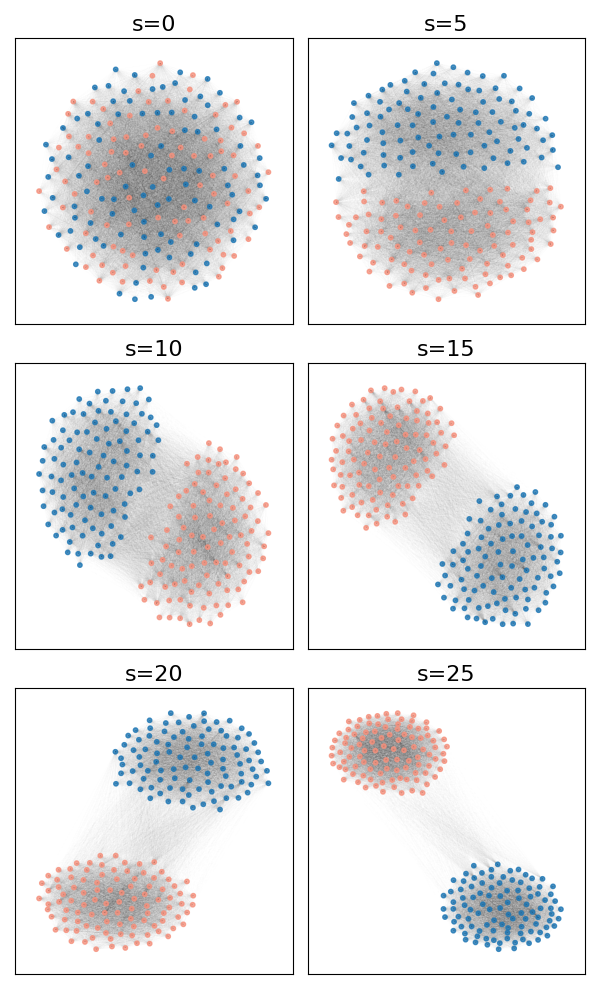}
  \caption{Illustration of the cases considered in Fig.~\ref{fig:diffcompnorm} in terms of Jaccard similarity networks obtained by using the respective parameters. Respectively to the results obtained for uniform densities, a more pronounced separation of the two groups can be observed for the same progression of $s$ values.}
  \label{fig:normal_s}
\end{figure}

Figure~\ref{fig:normal_s} depicts the results obtained for the normal densities in terms of respective Jaccard coincidence networks. Now, a more pronounce separation of the nodes can be observed as the parameters $s$ undergoes the same increase.

\section{Controlling the Sharpness of the Comparisons}
\label{sec:power}

The similarity values provided by the multiset Jaccard similarity index are intrinsically normalized in the interval $[0,1]$. This property allows the consideration of taking the power of that index~\cite{costafurther,costasimil}, as follows:
\begin{equation}\label{eq:defpow}
S(\p{x}, \p{y}) = J(\p{x}, \p{y})^D =
    \left(\frac{\sum_{i=1}^{n}\min(x_i^P, y_i^P)+\min(|x_i^N|,|y_i^N|)}
         {\sum_{i=1}^{n}\max(x_i^P, y_i^P)+\max(|x_i^N|,|y_i^N|)}\right)^D,
\end{equation}
where $D \in \mathbb{R}$ and $D > 0$. Again, we have that this index is bound within the interval $[0,1]$.

It can be verified~\cite{costafurther,costasimil,costamneurons} that sharper similarity comparisons will be obtained when larger values of $D$ and employed. Thus, it becomes possible to control the sharpness of the comparisons and to select a level for each specific application.

it is known that
\begin{equation*}
  f_\S(S)dS = f_\J(J)dJ
\end{equation*}
and therefore
\begin{equation}
  \label{eq:fs}
  f_\S(S) = \frac{1}{DS^{1-1/D}} f_\J(S^{1/D}).
\end{equation}

With this, given a distribution of values of \(J\) computed as in
section~\ref{sec:jaccard}, it is possible to directly compute the
distribution of \(S\). It is important to notice that there is a
problem with this approach for \(J\) close to zero, as the
approximations used in section~\ref{sec:jaccard} give a finite value
for the probability of \(J=0\) and larger than expected values for
\(J\) close to zero. This introduces problems when evaluating
Eq.~\eqref{eq:fs} for \(S\) close to zero.

For the distributions of Eq.~\eqref{eq:fxfyuni} the resulting expression is
\begin{multline}
  \label{eq:pdfdifferentD}
  f_\cS(S) = \frac{3}{\sqrt{2\pi}}\frac{1}{D S^{1-1/D}}\times\\
  \times \frac{224 n^2 - 50 n s - 61 s^2 + (160 n^2 + 234 n s - 88 s^2) S^{1/D}}
       {[32 n - 21 s + (16 n + 15 s) S^{1/D} + (80 n + 162 s) S^{2/D}]^{3/2}}\times\\
  \times \exp\left(-\frac{2 [-2 n + s + (10 n + 4 s) S^{1/D}]^2}
                  {32 n - 21 s + (16 n + 15 s) S^{1/D} + (80 n + 162 s) S^{2/D}}
      \right).
\end{multline}

\begin{figure}[t]
  \begin{subfigure}{0.45\textwidth}
    \includegraphics[width=\textwidth]{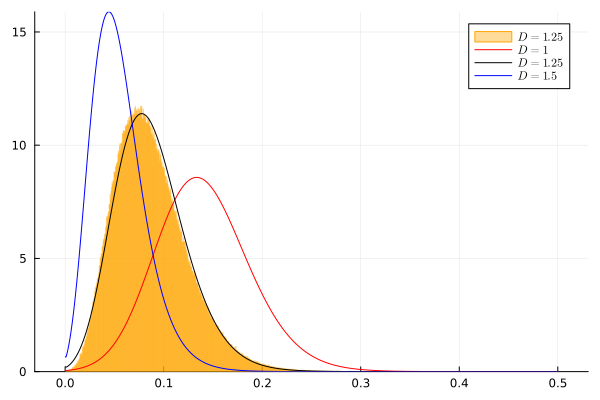}
    \caption{}
    \label{fig:pow}
  \end{subfigure}
  \begin{subfigure}{0.45\textwidth}
    \includegraphics[width=\textwidth]{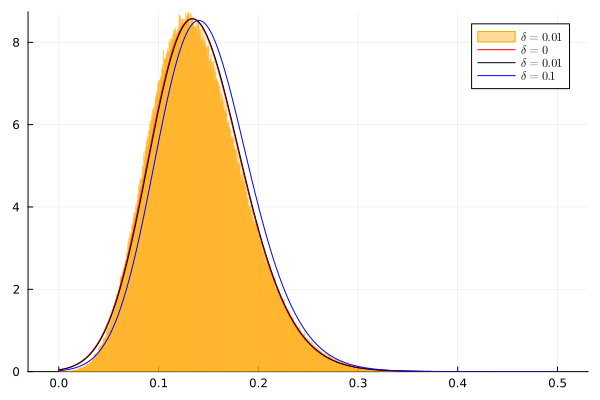}
    \caption{}
    \label{fig:reg}
  \end{subfigure}
  \caption{Effect of (a) an exponent and (b) regularization on the distribution of \(\mathcal{J}\). Lines correspond to Eq.~\eqref{eq:pdfdifferentD} and \eqref{eq:pdfdifferentdelta}, with \(n=25\) and \(s=10\); filled regions correspond to histograms of generated data.}
  \label{fig:powreg}
\end{figure}

Figure~\ref{fig:pow} shows the effect of the use of \(D\) and a comparison with generated data respectively to a reference similarity density. It can be verified that the adoption of increasing values of $D$ implies the similarity density to shift from right to left at the same time that its dispersion decreases. The former effect confirms the tendency of more strict comparisons being implemented for larger values of $D$, which is further illustrated in Figure~\ref{fig:powreg}.

\begin{figure}[t]
  \centering
  \includegraphics[width=0.9\textwidth]{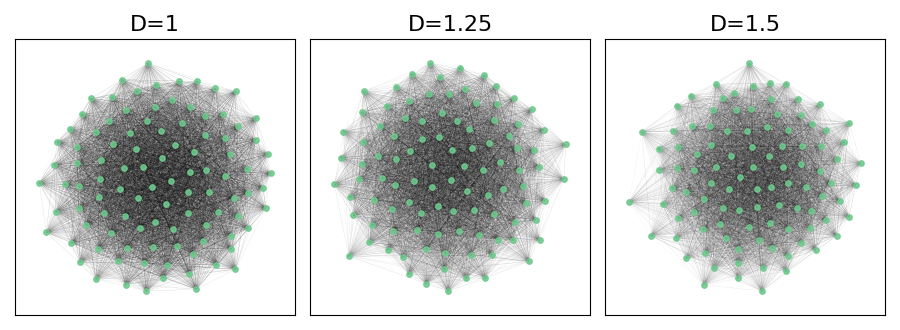}
  \caption{Illustration of the effect of the parameter $D$ on the Jaccard similarity values within a same group. A successive decrease of similarity values can be observed as $D$ increases, indicating that more strict comparisons are respectively being performed.}
  \label{fig:uniform_D}
\end{figure}

\section{Regularization}
\label{sec:regularization}

In its original form aimed at comparing sets, the Jaccard similarity index can lead to a singularity of the type $0/0$ when comparing two empty sets. Since this can also happen when applying the multiset Jaccard index for real valued vectors, it is interesting to consider means of addressing this singularity, which could also allow a smoother derivative around the comparison peak.

A simple possibility for \emph{regularizing} (or stabilizing) the Jaccard similarity index involves adding the same small constant to both its numerator and denominator. The regularized version of the Jaccard similarity, expressed as \(\hat{J}\), is given by
\begin{equation}
  \label{eq:regj}
  \hat J(\p{x}, \p{y}) =
  \frac{\sum_{i=1}^{n}\left[\min(x_i^P, y_i^P)+\min(|x_i^N|,|y_i^N|)\right]+\delta}
  {\sum_{i=1}^{n}\left[\max(x_i^P, y_i^P)+\max(|x_i^N|,|y_i^N|)\right]+\delta},
\end{equation}
where \(\delta\) is the arbitrary regularization factor.

With this, instead of Eq.~\eqref{eq:sum} the expression
\begin{multline}\label{eq:sumreg}
  \sum_{i=1}^{n} \left[\left(\min(x_i^P, y_i^P)+\min(|x_i^N|,|y_i^N|)\right)\right. -\\
  \left.-\hat{J} \left(\max(x_i^P,
      y_i^P)+\max(|x_i^N|,|y_i^N|)\right)\right]\le -(1-\hat{J})\delta
\end{multline}
results. Using again the approximations as in
Section~\ref{sec:jaccard}, the cumulative probability of \(\hat{J}\)
is approximated by the normal being less than \(-(1-\hat{J})\delta\),
which results in
\begin{equation}
  \label{eq:cdfnormreg}
  \frac{1}{2}\left[1+\erf\left(- \frac{\mu(\hat
        J)+(1-\hat{J})\delta}{\sqrt{2\sigma^2(\hat J)}}\right)\right],
\end{equation}
where \(\mu(\hat J)\) and \(\sigma^2(\hat J)\) are the same as in Eq.~\eqref{eq:mu} and \eqref{eq:sigma2}. From this the density is
\begin{equation}
  \label{eq:derivreg}
  f_{\hat\J}(\hat J)=\frac{(\mu+(1-\hat{J})\delta)\frac{d\sigma^2}{d\hat{J}}-2\left(\frac{d\mu}{d\hat{J}}-\delta\right)\sigma^2}{2\sqrt{2\pi}(\sigma^2)^{3/2}}\exp\left(-\frac{(\mu+(1-\hat{J})\delta)^2}{2\sigma^2}\right).
\end{equation}

For the special case of the distributions in Eq.~\eqref{eq:fxfyuni} these values
\begin{multline}
  \label{eq:pdfdifferentdelta}
  f_{\hat{\J}}(\hat J) = \frac{3}{\sqrt{2\pi}}\times\\
  \times \frac{224 n^2 - 50 n s - 61 s^2 + (640n - 216s)\delta + (160 n^2 + 234 n s - 88 s^2) J}
       {[32 n - 21 s + (16 n + 15 s) \hat{J} + (80 n + 162 s) \hat{J}^2]^{3/2}}\times\\
  \times \exp\left(-\frac{2 [-2 n + s -24\delta + (10 n + 4 s + 24\delta) J]^2}
                  {32 n - 21 s + (16 n + 15 s) \hat{J} + (80 n + 162 s) \hat{J}^2}
      \right).
\end{multline}

\begin{figure}[t]
  \centering
  \includegraphics[width=0.9\textwidth]{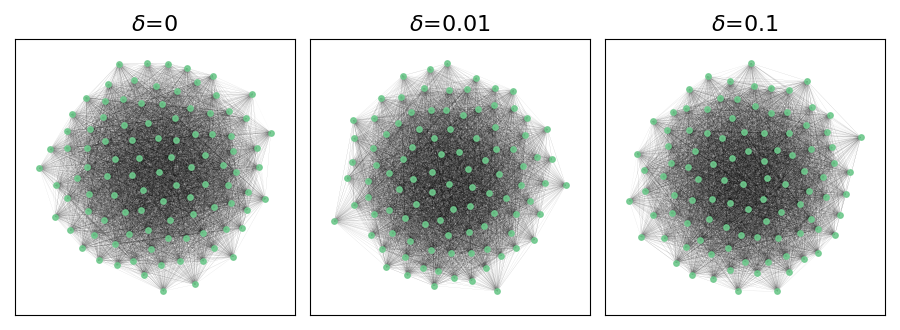}
  \caption{Jaccard similarity networks obtained for the same group of 100 points respectively to $\delta = 0, 0.01, 0.1$. The effect of the implemented regularizations cannot be visually identified.}
  \label{fig:uniform_delta}
\end{figure}

Figure~\ref{fig:reg} shows the effect of the regularization factor \(\delta\) on the distribution of values of similarity and includes a comparison with generated data. The obtained results indicate that larger values of $\delta$ tend to slightly shift the density to the right-hand side, therefore marginally decreasing the sharpness of the comparisons which leads to an increase of the obtained similarities. Nevertheless, as further illustrated in Figure~\ref{fig:uniform_delta}, the effect is very slight, making it possible to avoid the singularity problem with much distortion in the comparison.

\section{Concluding remarks}
\label{sec:conclusion}

Comparing entities represented in terms of their respective features constitutes a central operation frequently needed in science and technology, as it underlies a large number of related approaches including data visualization, analysis, classification, and modeling.

In the present work, an analytical statistical approach has been developed for estimating the distribution of the values obtained by the multiset Jaccard similarity index applied to multivariate features characterized by respective statistical densities.

By assuming the several involved features to be statistically independent as well as having sufficiently large dimensionality, it has been possible to apply the central limit theorem in order to obtain expressions for estimation of the mean and variance of the Jaccard similarity values, which can then be employed in order to obtain a generic expression of Jaccard index distribution.

Illustrations of the developed approach were reported respectively to the uniform and normal densities, characterized by good agreement between the analytical and numeric estimations.

In addition, modifications of the Jaccard similarity index so as to control the sharpness of the implemented comparison, as well as in order to account for some regularization of this index, have also been analytically addressed.

The reported concepts, expressions, and results motivate many possible further developments. Results for different statistical distributions, varying the number of features, and considering more than two groups can be easily done with the same approach. An open question is how to deal with low dimensionality and correlations among the features. Another interesting prospect corresponds to extending the described developments in order to obtain analytical statistical expressions which could be employed to complement the characterization of the coincidence similarity index.

\section*{Acknowledgments}
A. Benatti is grateful to MCTI PPI-SOFTEX (TIC 13 DOU 01245.010\\222/2022-44), FAPESP (grant 2022/15304-4), and CNPq. Luciano da F. Costa thanks CNPq (grant no.~307085/2018-0) and FAPESP (grant 2022/15304-4).

\bibliography{ref}
\bibliographystyle{unsrt}

\end{document}